\documentclass[aps,prl,showpacs,amsmath,amssymb,superscriptaddress,reprint,10pt]{revtex4-1}
\usepackage{bm}
\usepackage{graphicx}
\usepackage{xcolor}
\usepackage{url}
\usepackage[normalem]{ulem}
\usepackage{verbatim}
\usepackage{bbm}
\usepackage{subfigure}
\usepackage{mathrsfs}
\usepackage[colorlinks=true,breaklinks=true,allcolors=blue]{hyperref}

\definecolor{david}{rgb}{1,0,0}

\definecolor{julien}{rgb}{0,.8,.5}

\newcommand{\dep}[2]{\dfrac{\partial #1}{\partial #2}}

\renewcommand{\d}{\mathrm{d}}

\renewcommand{\Re}{\operatorname{Re}}
\renewcommand{\Im}{\operatorname{Im}}

\renewcommand{\L}{\operatorname{\mathscr{L}}}
\newcommand{\N}{\operatorname{\mathscr{N}}}

\begin{document}

\title{Bifurcations and singularities for coupled oscillators with inertia and frustration}  

\author{J. Barr\'e}
\affiliation{Laboratoire J.-A. Dieudonn\'e, Universit\'e de Nice Sophia-Antipolis, UMR CNRS 7351, Parc Valrose 06108 Nice Cedex 02 France, and Institut Universitaire de France, 75005 Paris, France.\\}
\author{D. M\'etivier}
\affiliation{Laboratoire J.-A. Dieudonn\'e, Universit\'e de Nice Sophia-Antipolis, UMR CNRS 7351, Parc Valrose 06108 Nice Cedex 02, France.}

\date{\today}

\begin{abstract}
We prove that any non zero inertia, however small, is able to change the nature of the synchronization transition in Kuramoto-like models, either from continuous to discontinuous, or from discontinuous to continuous. This result is obtained through an unstable manifold expansion in the spirit of J.D. Crawford, which features singularities in the vicinity of the bifurcation. Far from being unwanted artifacts, these singularities actually control the qualitative behavior of the system. Our numerical tests fully support this picture. 
\end{abstract}

\maketitle 

\begin{center}
  \begin{figure}
     \includegraphics[width=0.48\textwidth]{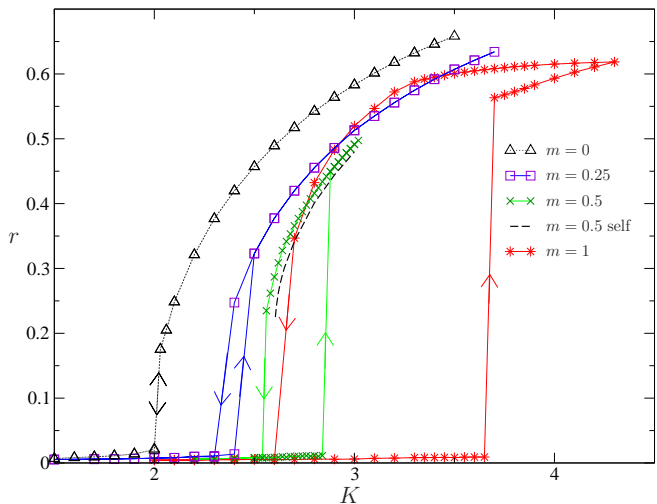}
    \caption{Asymptotic order parameter $r$ as a function of $K$ for different $m$ ($\alpha=0$). The arrows indicate the direction of the jumps. Without inertia, the transition is continuous, while an hysteresis appears already for small $m$. Note the presence of a single branch with $r\neq 0$ for $m=0.25,~0.5$, while there are two for $m=1$. The dashed line is the partially synchronized solution given by the self-consistent method (see \cite{suppl}) for $m=0.5$. The frequency distribution is Lorentzian: $g_\sigma(\omega)=(\sigma/\pi)/(\sigma^2+\omega^2)$ with $\sigma=1$.} 
    \label{fig:m_change}
  \end{figure}
\end{center}

Understanding synchronization in large populations of coupled oscillators is a question which arises in many different fields, from physics to neuroscience, chemistry and biology~\cite{Pikovsky}. Describing the oscillators with their phases only, Winfree~\cite{Winfree}, and Kuramoto~\cite{Kuramoto75} have introduced simple models for this phenomenon. The latter model, which features a sinusoidal coupling, and an all-to-all interaction between oscillators, has become a paradigmatic model for 
synchronization, and its very rich behavior prompted an enormous number of studies. Kuramoto model displays a transition between an incoherent state, where each oscillator rotates at its own 
intrinsic frequency, and a state where at least some oscillators are phase-locked. The degree of coherence is measured by an order parameter $r$, which bifurcates -continuously for symmetric unimodal frequency distributions- from $0$ when the coupling is increased, or the dispersion in intrinsic frequencies decreases.
In order to better fit modeling needs, it has been necessary to consider refined models, including for instance, citing just a few contributions: more general coupling~\cite{Daido92}, noise~\cite{Sakaguchi88}, phase shifts bringing frustration~\cite{Sakaguchi88b}, delays~\cite{Izhikevich98,Yeung99}, or a more realistic interaction topology~\cite{Sakaguchi87,Hong02}. 
In particular, inertia has been introduced to describe the synchronization of a certain firefly~\cite{Ermentrout91}, and proved later useful to 
model coupled Josephson junctions~\cite{Wiesenfeld96,Trees05} and power grids~\cite{Filatrella08,Olmi14}; recently, an inertial model on a complex network was shown
to display a new type of "explosive synchronization"~\cite{Ji13}.
It has been quickly recognized~\cite{Tanaka97,Acebron98} that inertia could turn the continuous Kuramoto transition into a discontinuous one with hysteresis. At first
sight, a natural adaptation of the original clever self-consistent mean-field approach by Kuramoto~\cite{Kuramoto75} seems to 
explain satisfactorily this observation~\cite{Tanaka97,Tanaka97b}: a sufficiently large inertia induces a bistable dynamical behavior of some oscillators, that
translates into a hysteretic dynamics at the collective level.  
However, Fig.~\ref{fig:m_change} makes clear that even a small inertia is enough to trigger a discontinuous transition: this cannot be accounted for by the bistability picture. 

In this letter, we explain why any non zero inertia, however small, can have a dramatic effect on the transition: it can turn discontinuous an otherwise continuous transition,  \emph{and the other way round}. These results are obtained through a careful unstable manifold expansion in the spirit of \cite{Crawford_Vlasov,Crawford_Kura,Crawford_Kura_prl} (see also \cite{Strogatz_review} for a very readable discussion of the method), which uses the instability rate of the incoherent state as a small parameter: in the absence of noise, the linearized problem features a continuous spectrum on the imaginary axis, which may induce singularities in the usual expansions. We point out that these singularities related to the continuous spectrum are key for a comprehensive understanding of the bifurcations: they
control the phase diagram in presence of frustration, as well as the Hamiltonian limit, where the very strong nonlinear effects of Vlasov equation are recovered.
We compare our predictions with large-scale numerical simulations: using a GPU (graphics processing unit) architecture allows us to reach a number of oscillators significantly larger than in most previous works; this is crucial to test with a reasonable precision scaling laws in the vicinity of bifurcations. 

We believe that these results should establish the singular expansions "\`a la Crawford" as another method of choice to understand the qualitative behavior of 
Kuramoto-like models, along the original mean-field self-consistent method, the Ott-Antonsen ansatz \cite{Ott08,Martens09}, and the bifurcation methods used in \cite{Acebron98,Acebron00,Acebron05,Komarov14}. Indeed this method is applicable for generic distribution function and interaction, and provides information on the order of the transition and scaling laws close to the bifurcation.

{\em The model.---} Our starting point is the model introduced by Tanaka et al. \cite{Tanaka97}, which adds inertia to the original Kuramoto model.  It has been  since then studied by many authors, often in presence of noise, and we first discuss some of the theoretical results obtained so far.
\cite{Tanaka97} adapts the original self-consistent Kuramoto method to the presence of inertia, and predicts, consistently with the numerics, that
a large enough inertia makes the transition discontinuous. The small inertia case was apparently not studied. In \cite{Acebron98,Acebron00}, the authors perform a bifurcation study of the incoherent state in presence of noise, and find a critical inertia beyond which the transition should be discontinuous; their result suggests that a small inertia \emph{can} make a qualitative difference, but the singular nature of the small-noise limit makes an extrapolation to zero noise difficult.
We note that a full "phase diagram" compatible with \cite{Acebron98,Acebron00} is presented in \cite{Gupta14} (see also \cite{Komarov14}). In the following, we also add to the model in \cite{Tanaka97} a "frustration" parameter $\alpha$, as in \cite{Sakaguchi88b}; this will provide us with a further parameter to make testable predictions. Our resulting model is then the same as \cite{Komarov14}, without noise.

Each of the $N$ oscillators in the system has a frequency $v_i$, with $i\in {1,\cdots,N}$ and a phase $\theta_i\in [0,2\pi[$; it also has a natural frequency $\omega_i$, 
drawn from a frequency distribution $g$. We assume that $g$ is even ($g(-\omega)=g(\omega)$).
If there is no coupling between oscillators, the actual frequency  $v_i$ tends to the natural frequency $\omega_i$. The dynamical equations for positions and velocities are
\begin{subequations}
\begin{align}
\dot{\theta_i} &=v_i 
\\m\dot{v_i}&=\gamma(\omega_i-v_i)+\dfrac{K}{N}\sum_{j=1}^N\sin(\theta_j-\theta_i-\alpha).
\end{align}\label{eq:particles}
\end{subequations}
Note that the above notations are slightly different from Tanaka's, to make the Kuramoto and Vlasov limits easier to understand. The connection with the standard notations is $m_{\mbox{Tanaka}}=m/\gamma$, $K_{\mbox{Tanaka}}=K/\gamma$.
If the inertia $m$ tends to $0$, one recovers the usual Kuramoto model after a suitable change in the time variable,
\begin{equation}
\dot{\theta_i}=\omega_i+\dfrac{K}{N}\sum_{j=1}^N\sin(\theta_j-\theta_i-\alpha)
\end{equation}
If $\gamma=0$, there is no restoring force towards the natural frequency, and one obtains for $\alpha=0$ a Hamiltonian model with an all-to-all coupling and a cosine interaction potential. It is usually called  HMF model in the literature, and has served as a simple paradigmatic model for mean-field Hamiltonian dynamics, see \cite{Campa09} for a review.
In the $N\to\infty$ limit, the system \eqref{eq:particles} is described by a kinetic equation for the phase space density $F(\theta,v,\omega,t)$: 
\begin{equation}
\begin{split}
\dep{F}{t}+v\dep{F}{\theta}+&\dfrac{K}{2i m}\left (r_1[F] e^{-i\theta}e^{-i\alpha}-r_{-1}[F] e^{i\theta}e^{i\alpha}\right )\dep{F}{v}
\\&-\dfrac{\gamma}{m}\dep{}{v}\left ((v-\omega)F\right)=0,
\label{eq:pde}
\end{split}
\end{equation}
where the usual order parameter is $r=|r_1|$, with
\begin{equation}
r_k[F]=\int F(\theta,v,\omega,t) e^{ik\theta} ~\d \theta\,\d v \, \d \omega.
\end{equation}

{\em Unstable manifold expansion.---} The incoherent stationary solution corresponds to each oscillator running at its natural frequency, with phases homogeneously distributed: $F(\theta,v,\omega,t)=f^0(v,\omega) = g(\omega) \delta(v-\omega)/(2\pi)$. It is easy to check that $f^0$ is indeed a stationary solution of \eqref{eq:pde}. 
Increasing the coupling strength $K$, $f^0$ changes from stable to unstable. Our goal is to study the dynamics of \eqref{eq:pde} in the vicinity of this 
bifurcation.

For this purpose, we first decompose eq.\eqref{eq:pde} in a linear and a nonlinear part, with $F=f^0+f$:
\begin{equation}
\dep{f}{t}=\L f+\N [f],
\label{VKdec}
\end{equation}
with
\begin{equation}
\begin{split}
\L f=-v\dep{f}{\theta}-\dfrac{K}{2i m}&\left (r_1[f] e^{-i\theta}e^{-i\alpha}-r_{-1}[f] e^{i\theta}e^{i\alpha}\right )\partial_v f^0
\\&~~~~+\dfrac{\gamma}{m}\dep{}{v}\left ((v-\omega)f\right ),
\label{eq:pde_lin}
\end{split}
\end{equation}
and 
\begin{equation}
\N [f]=-\dfrac{K}{2i m}\left (r_1[f] e^{-i\theta}e^{-i\alpha}-r_{-1}[f] e^{i\theta}e^{i\alpha}\right )\partial_v f.
\label{eq:pde_nonlin}
\end{equation}
The precise study of the linear operator $\L$ is an important building block in our non linear analysis, and we collect below the main results concerning $\L$ (details are given in the supplemental material \cite{suppl}). \eqref{eq:pde} is symmetric with respect to rotations
$(\theta,v,\omega)=(\theta+\varphi,v,\omega)$;  if $\alpha=0$ and $g(\omega)$ even, it is in addition symmetric with respect to reflections
$(\theta,v,\omega)=-(\theta,v,\omega)$~\cite{Crawford_Kura}. In this article we take $g$ even, and we restrict to the case of two unstable eigenvectors. This is generically the case when: i) $\alpha\neq 0$; in this case there is a complex unstable eigenvalue $\lambda$, and $\lambda^\star$ is also an unstable eigenvalue; 
ii) $\alpha=0$, and $\lambda$ is real; in this case it is twice degenerate, associated with two eigenvectors. Hence in both cases we will build a two-dimensionnal unstable manifold. We leave for future studies the cases $\alpha=0$, $\lambda$ complex, which leads to a four-dimensionnal unstable manifold \cite{Crawford_Kura}, as well as non even $g(\omega)$ distributions. 
$\L$ is diagonal when expressed in the Fourier basis for the phases. It is then easy to see that the discrete spectrum of $\L$ is associated with the $k=\pm 1$ Fourier modes, that is eigenvectors are proportional to $e^{\pm i\theta}$.  

$\Psi$, eigenvector of $\L$ associated with $\lambda$ and $k=1$, is given by
$\Psi(\theta,v,\omega)=\psi(v,\omega) e^{i \theta}$, with $\psi(v,\omega)=U_0(\omega)\delta(v-\omega)+U_1(\omega)\delta'(v-\omega)$, and $\delta'$ stands for the first derivative with respect to $v$ of the Dirac distribution. The expression for $U_0$ and $U_1$ is provided in the supplemental material \cite{suppl}.
The dispersion relation, from which $\lambda$ is computed, reads:
\begin{equation}
\Lambda(\lambda)=1-\dfrac{Ke^{i\alpha}}{2m} \int\dfrac{g(\omega)}{(\lambda+\gamma/m+i\omega)(\lambda+i\omega)}\,\d \omega =0 .
\label{eq:dis_rel}
\end{equation}
This dispersion relation can be recovered as the noiseless limit of the one in \cite{Acebron00}, as it should. One can also check that the $\gamma\to 0$ limit 
yields the Vlasov dispersion relation with a cosine potential and $g(\omega)$ as stationary velocity profile; the $m\to0$ limit yields the standard Kuramoto dispersion relation.

We will need the projection $\Pi$ over the unstable eigenspace $\mathcal{V}=\mbox{Span}(\Psi,\Psi^\star)$. For this purpose, we
introduce the adjoint operator $\L^\dagger$, defined through $(f_1,\L f_2)=(\L^\dagger f_1,f_2)$, where 
$(f_1, f_2)=\iint f_1^\star f_2\,\d\omega\,\d v \,\d \theta$.
The adjoint eigenvector associated with the eigenvalue $\lambda$ is 
$ \tilde{\Psi}(\theta,v,\omega)=\tilde{\psi}(v,\omega)e^{i \theta}/2\pi$.
 We do not know how to compute $\tilde{\psi}(v,\omega)$ in closed form. However, in the following computations, $\tilde{\psi}$ only appears in scalar products with delta functions $\delta(v-\omega)$ and their derivatives; as a consequence, 
we only need to know $\tilde{\psi}^{(n)}(\omega) := \partial_v^{n} \tilde{\psi}(\omega,\omega)$. The expression for
$\tilde{\psi}^{(n)}(\omega)$ is provided in \cite{suppl}. Then
$\Pi \cdot \phi = \left(\tilde{\Psi},\phi \right) \Psi +\left(\tilde{\Psi}^\star,\phi \right) \Psi^\star$.
With this knowledge of the linear part $\L$, we now proceed to the non linear analysis. Following \cite{Crawford_Vlasov}, we introduce the unstable manifold
$\mathcal{M}$ associated to the stationary solution $f^0$. 
$\mathcal{M}$ is the set of functions $F$ that tend to $f^0$ when $t \to -\infty$. This is a manifold which dimension is the same as the linear unstable subspace $\mathcal{V}$, and it is clearly invariant by the dynamics. The tangent space to $\mathcal{M}$ at $f^0$ is
$\mathcal{V}$. Any element $\phi$ of $\mathcal{M}$ in a neighborhood of $f^0$ can be written as
\begin{equation}
\phi = A \Psi +A^\star \Psi^\star + H[A,A^\star](\theta,v,\omega).
\label{eq:manifold}
\end{equation}
$A \Psi +A^\star \Psi^\star$ is the projection of $\phi$ on $\mathcal{V}$ according to $\Pi$; hence $(\tilde{\Psi},H)=0$. Furthermore, $H=O((A,A^\star)^2)$. 
Note that this parametrization of the unstable manifold as a function of $A$ and $A^\star$ is valid close to $f^0$; however, it may not be globally valid.
For an initial condition on $\mathcal{M}$, and assuming that the dynamics remain in a region where the unstable manifold can be parametrized as in \eqref{eq:manifold}, the whole dynamics is parameterized by the function $A(t)$, which is related to $r$ by $r=2\pi |A| +O(|A|^3)$. Our goal is then to determine the evolution equation for $A$.
$H$ itself is of course unknown and has to be determined at the same time as the dynamical equation for $A$. 
The strategy is to build an expansion in $A$:
\begin{eqnarray}
\frac{dA}{dt} &=& \lambda A +c_3 |A|^2 A + O(A^5)  \label{eq:reduced}\\
H(A,A^\star) &=& A^2 \mathcal{H}_2^{20}(\theta,v,\omega) +AA^\star \mathcal{H}_2^{11}(\theta,v,\omega)+\ldots \nonumber
\end{eqnarray}
Note that the $A \leftrightarrow -A$ symmetry is responsible for the particular form of the cubic term and the vanishing of the quadratic terms. Using the linear order for $dA/dt$, it is possible to compute $H$ at quadratic order in $(A,A^\star)$ by solving linear equations. $H$ at quadratic order then gives access to $c_3$. The resulting expression for $c_3$ is easy to analyze in the interesting limit $\lambda_R =\Re(\lambda) \to 0^+$.
At the expense of increasingly intricate computations, one could go on with this scheme; we have stopped at $c_3$. We give below the key results of the computation, whereas all details are presented in \cite{suppl}.

\begin{center}
  \begin{figure}
     \includegraphics[width=0.48\textwidth]{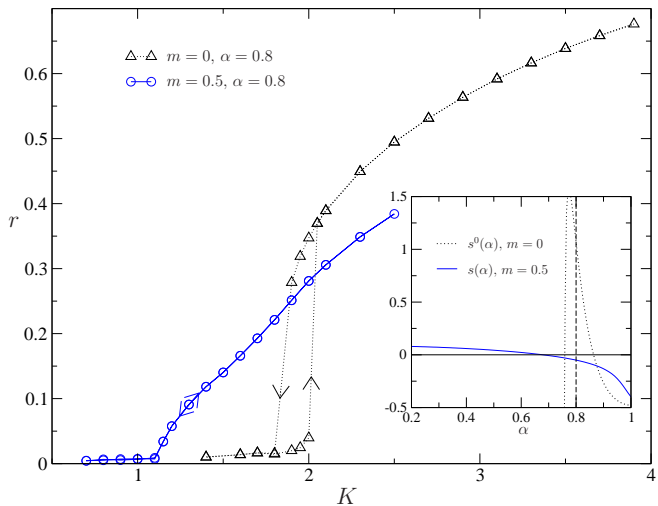}
    \caption{Asymptotic $r$ as a function of $K$ for $\alpha=0.8$, with $m=0$ or $m=0.5$. The frequency distribution is a superposition of two Lorentzians as in \cite{Omel12}, Fig.3: $g(\omega)=\tau g_1+ (1-\tau)g_\delta$, with $\tau=0.8$, $\delta=0.075$; $g$ is unimodal.
 The inset shows that $s^0(\alpha=0.8)>0$ (hence discontinuous transition at $m=0$), and $s(\alpha=0.8)<0$ (hence continuous transition as soon as $m>0$).}
    \label{fig:superlo}
  \end{figure}
\end{center}

{\em Discussion.---} Using the reduced dynamics \eqref{eq:reduced} truncated at order $A^3$ provides essential qualitative informations: i)The bifurcation is subcritical 
(i.e. with a jump in the order parameter) if and only if $\Re\,(c_3)>0$ ii)In the supercritical case, one obtains the asymptotic order parameter 
$|A|_\infty = \sqrt{-\lambda_R/\Re\,(c_3)}$. We have to evaluate $c_3$ close to the bifurcation point, that is when $\lambda_R \to 0$. We find, for $m>0$, with $\lambda_I=\Im(\lambda)$ (our hypothesis of a two-dimensional unstable manifold ensures that $\Lambda'(i\lambda_I)\neq 0$):
\begin{equation}
\Re\,c_3 \sim \frac{\pi^3}{2}\frac{mK^3}{\gamma^4}\frac{g(\lambda_I)}{\lambda_R} \Re\left (\dfrac{e^{i\alpha}}{\Lambda'(i\lambda_I)}\right ).
\label{eq:c3}
\end{equation}
From this, the dramatic effect of the inertia $m$ appears clearly: it introduces into $c_3$ a contribution diverging like $1/\lambda_R$, which is the dominant 
one: the sign of $s=\Re\,(e^{i\alpha}/\Lambda'(i\lambda_i))$ controls the bifurcation type, sub-(resp. super) critical for $s>0$ (resp. $s<0$). For $m=0$, the next order term, which does not diverge when $\lambda_R\to 0$, is needed; the bifurcation is then controlled by 
$s^0={\rm Re}(\Lambda''(i\lambda_I)/\Lambda'(i\lambda_I))$: sub-(resp. super) critical for $s_0>0$ (resp. $s_0<0$) (this generalizes to $\alpha\neq 0$ a result of \cite{Crawford_Kura_prl}, see \cite{suppl}.)
Hence, any small $m$ may either turn a supercritical bifurcation at $m=0$ into a subcritical one,
or \emph{the other way round, turn a subcritical bifurcation at $m=0$ into a supercritical one}. While the first direction, illustrated on Fig.\ref{fig:m_change}, was anticipated in \cite{Gupta14,Komarov14}, the second direction is unexpected. Fig.\ref{fig:superlo} provides an example. 
Furthermore, in the supercritical case, we predict the scaling law for the asymptotic order parameter $|A|_\infty \propto \lambda_R$, and this is also observed. If the distribution $g$ is unimodal we note that $s(\alpha =0)>0$, so the bifurcation is always subcritical.
Finally, \eqref{eq:c3} makes clear that both the standard first order Kuramoto ($m=0, \alpha=0$) and Vlasov ($\gamma=0$) limits are singular. In the first case, the divergent term vanishes, and the bifurcation is controlled by the sign of $s^0$. One recovers the already known results: for a symmetric unimodal $g$, $s^0(\alpha=0)<0$ and the bifurcation is supercritical, with standard scaling
 $|A|_\infty \propto \sqrt{\lambda_R}$. In the second case, \eqref{eq:c3} diverges when $\gamma \to 0^+$. Redoing the computations in the limit $\gamma\to 0$ indeed yields (for $\alpha=0$) $c_3\propto-\frac{1}{\lambda_R^3}$, as found in \cite{Crawford_Vlasov}. This leads to the "trapping scaling" well known in plasma physics
$|A|_\infty \propto \lambda_R^2$. 
{\em Numerics.---} We present in this section precise numerical simulations that fully support the above analysis. The time-evolved system is obtained via GPU parallel
implementation of a Runge-Kunta 4 scheme for the equations \eqref{VKdec} \cite{Gupta14}. The order parameter is computed by its standard discrete definition~\cite{Kuramoto75}. For every simulation we take $N=65536$, $\gamma=1$, and a time step $\Delta t=10^{-3}$. The asymptotic order parameter $r$ is the average of $|r_1|(t)$ for $t\in [1500,2000]$. In order to test our prediction on the type of bifurcation, we start from an unsynchronised state (drawing positions $\theta_i$ uniformly on a unit circle). The $\omega_i$ are sampled according to $g$, the initial velocities are $v_i=\omega_i$. 
We let the system evolve until $t=2000$ and measure the averaged order parameter. Then we vary the coupling constant $K\to K+\Delta K$ with $\Delta K=0.1$ or $0.2$ (or smaller close to transitions) and reiterate the procedure; at some point the bifurcation towards synchronisation is observed. When $K$ is large enough we apply the same procedure in the other direction, $K\to K-\Delta K$. Thus, we are able to distinguish clearly a subcritical bifurcation (with a characteristic hysteresis cycle) from a supercritical bifurcation (with no hysteresis).  On Fig.\ref{fig:m_change}, we see how the hysteretic cycle depends on the inertia $m$. For $m=1$, there are two branches with $r\neq 0$: these correspond to the bistable behavior of the single oscillator dynamics in a range of 
 $\omega$, see~\cite{Tanaka97}; for $m=0.5$ and $m=0.25$, the single oscillator dynamics is not bistable in the transition region, and, accordingly, there is only one branch with $r\neq 0$. The bifurcation remains nevertheless clearly subcritical. On Fig.\ref{fig:superlo}, inertia induces a supercritical
transition; \eqref{eq:c3} also correctly predicts the linear scaling of the saturated state in this case.
Finally, we note that in the subcritical regime, the numerically observed $K_c$ is sometimes lower than the prediction \eqref{eq:pde}; this is presumably related to strong finite size effects \cite{Hong07}, especially in presence of inertia\cite{Olmi14}.

{\em Conclusions.---} 
We have constructed an unstable manifold expansion for models of synchronization with inertia and frustration, circumventing the problem of the continuous spectrum on the imaginary axis.
The singularities appearing in the expansion may at first sight seem harmful, but they actually control the system's behavior in the vicinity of the bifurcation, and allow useful qualitative and quantitative predictions. In particular, while synchronization models tend to present complicated phase diagrams for which it is difficult to develop an intuition \cite{Omel12,Komarov14}, we obtain simple criteria determining the character of the transition.
We note that, since the unstable manifold is not attractive, the reduced description could be valid only for specific initial conditions; numerics show that its validity is much wider than what might have been expected. The versatility of the method suggests that it can be adapted to many different situations. Finally, we remark that the bifurcation diagram of the standard Kuramoto model (in particular, without inertia) has been established rigorously very recently \cite{Chiba,Dietert}. It is tempting to relate this mathematical success to the absence of singularities in the corresponding unstable manifold expansion, although the exact relationship between these facts is still unknown.

\emph{Acknowledgments}: Part of this work has been performed while J.B. was on leave at Imperial College and University of Orl\'eans. The authors acknowledge useful discussions with S. Gupta, S. Olmi and A. Torcini.

\appendix

\section{Unstable manifold expansion}

\subsection{Notations}
Our basic equation is kinetic equation for the phase space density $F(\theta,v,\omega,t)$:
\begin{equation}
\begin{split}
\dep{F}{t}+v\dep{F}{\theta}+&\dfrac{K}{2i m}\left (r_1[F] e^{-i\theta}e^{-i\alpha}-r_{-1}[F] e^{i\theta}e^{i\alpha}\right )\dep{F}{v}
\\&-\dfrac{\gamma}{m}\dep{}{v}\left ((v-\omega)F\right)=0,
\label{eq:pde}
\end{split}
\end{equation}
where
\begin{equation}
r_k[F]=\int F(\theta,v,\omega,t) e^{ik\theta} ~\d \theta\,\d v \, \d \omega.
\end{equation}
The incoherent stationary solution is $F(\theta,v,\omega,t)=f^0(v,\omega) = g(\omega) \delta(v-\omega)/(2\pi)$; it corresponds to the situation where the velocity of each oscillator is equal to its natural frequency, and the phases are homogeneously distributed in $[0,2\pi[$. 
We write Eq.\eqref{eq:pde} as a sum of a linear and a nonlinear parts, with $F=f^0+f$:
\begin{equation}
\dep{f}{t}=\L f+\N [f],
\label{VKdec}
\end{equation}
with
\begin{equation}
\begin{split}
\L f=-v\dep{f}{\theta}-\dfrac{K}{2i m}&\left (r_1[f] e^{-i\theta}e^{-i\alpha}-r_{-1}[f] e^{i\theta}e^{i\alpha}\right )\partial_v f^0
\\&~~~~+\dfrac{\gamma}{m}\dep{}{v}\left ((v-\omega)f\right ),
\label{eq:pde_lin}
\end{split}
\end{equation}
and 
\begin{equation}
\N [f]=-\dfrac{K}{2i m}\left (r_1[f] e^{-i\theta}e^{-i\alpha}-r_{-1}[f] e^{i\theta}e^{i\alpha}\right )\partial_v f.
\label{eq:pde_nonlin}
\end{equation}

\subsection{Study of the linear part} 
We first study the symmetries of \eqref{eq:pde}, which give informations on the degeneracy of the eigenvalues. They will also constrain the structure of the expansion later on. \eqref{eq:pde} is symmetric with respect to rotations $R_\varphi(\theta,v,\omega)=(\theta+\varphi,v,\omega)$;  
if $\alpha=0$ and $g(\omega)$ even, it is in addition symmetric with respect to reflections $\kappa(\theta,v,\omega)=-(\theta,v,\omega)$. 
Notice that $\L$ is diagonal when expressed in the Fourier basis for the phases, that is for $\Psi(\theta,v,\omega) =\sum_k \Psi_k(v,\omega)e^{ik\theta}$, we have
\[
\L \Psi = \sum_k (\L_k \Psi_k)e^{ik\theta}
\]
where $\L_k$ acts over a function $h(v,\omega)$ as
\begin{equation}
\begin{split}
\L_k h = & -ikv h -\dfrac{K}{2i m} 2\pi \int h \,\d v\,\d\omega \left ( e^{-i\alpha}\delta_{k,-1}-e^{i\alpha}\delta_{k,1}\right )\partial_v f^0  \\
&+\dfrac{\gamma}{m}\dep{}{v}\left[ (v-\omega)h\right] 
\end{split}
\end{equation}
We may then study separately the spectrum of each operator $\L_k$. We are interested in the point spectrum (i.e. eigenvalues), which may exist only for 
$k=\pm 1$. Let us solve $\L_1\psi = \lambda \psi$. This yields
\[
(\lambda+iv)\psi = \frac{\gamma}{m} \dep{}{v}\left ((v-\omega)\psi\right ) +\dfrac{K}{2i m}e^{i\alpha} \int \psi \,\d v\,\d\omega g(\omega)\delta'(v-\omega).
\]
Looking for a solution in the form 
\begin{equation}
\psi = U_0(\omega)\delta(v-\omega) +U_1(\omega)\delta'(v-\omega)
\label{eq:varpsi}
\end{equation}
and imposing the normalisation $\int \psi \,\d v\,\d\omega=1$, one finds
\begin{eqnarray}
U_0 &=& \frac{K}{2m}e^{i\alpha}\frac{g(\omega)}{(\lambda+i\omega)(\lambda +\gamma/m +i\omega)} \\
U_1 &=& \frac{K}{2im}e^{i\alpha}\frac{g(\omega)}{\lambda +\gamma/m +i\omega}.
\end{eqnarray}
Expliciting the normalisation condition yields the dispersion relation:
\begin{equation}
\Lambda_1(\lambda) = 1-\frac{K}{2m}e^{i\alpha} \int \frac{g(\omega)}{(\lambda+i\omega)(\lambda +\gamma/m +i\omega)}\,\d\omega =0
\label{eq:Lambda}
\end{equation}
where we have introduced the "spectral functions" $\Lambda_k$. 
The same computation can be performed using $\L_{-1}$. One obtains a similar spectral function:
\begin{equation}
\Lambda_{-1}(\lambda) = 1-\frac{K}{2m}e^{-i\alpha} \int \frac{g(\omega)}{(\lambda-i\omega)(\lambda +\gamma/m -i\omega)}\,\d\omega =0.
\label{eq:Lambda-1}
\end{equation}
Hence, if $\lambda$ is an eigenvalue of $\L_1$, $\lambda^\ast$ is an eigenvalue of $\L_{-1}$, as expected from the rotation symmetry.
It is also easily seen that if $\alpha=0$ and $g$ even, $\Lambda_1(\lambda^\ast)=\Lambda_1(\lambda)^\ast$. Hence if $\lambda$ is an eigenvalue
of $\L_1$, $\lambda^\ast$ is also an eigenvalue of $\L_1$. Of course, $\lambda$ and $\lambda^\ast$ are also eigenvalues of $\L_{-1}$. 
Thus, if there is an unstable eigenvalue with non zero imaginary part, its multiplicity is two, and there are four unstable eigenvectors: the unstable manifold 
is of dimension four. This case goes beyond the purpose of this paper, so that we restrict ourselves to the two following situations: \\
i) $\alpha \neq 0$\\
ii) $\alpha=0$, and the unstable eigenvalue is real.\\
In both cases, we also assume that $\Lambda'_{\pm 1}(\lambda)\neq 0$ (which is the generic situation), and there are only two unstable eigenvectors. The remaining situation where $\alpha=0$ and there is a complex unstable eigenvalue can happen,
for instance for a bimodal distribution $g$.

From now on we call $\Lambda(\lambda)=\Lambda_1(\lambda)$ the spectral function, forgetting the subscript $1$. 
We have two eigenvectors 
\begin{equation}
\Psi(\theta,v,\omega) =\psi(v,\omega) e^{i\theta}~\mbox{and}~\Psi^\ast(\theta,v,\omega) =\psi(v,\omega)^\ast e^{-i\theta}~,
\end{equation}
where $\psi$ is given by \eqref{eq:varpsi}.

To define the projection over $\mbox{Span}(\Psi,\Psi^\ast)$, we need to study $\L^\dagger$, the adjunct of $\L$.
We define $\L^\dagger$ through the scalar product
\begin{equation}
(f_1, f_2) =\int\left <f_1,f_2\right >\,\d \theta=\int\int f_1^\ast f_2\,\d\omega\,\d v \,\d \theta.
\end{equation}
We obtain
\begin{equation}
\begin{split}
\L^\dagger f_1=&v\partial_\theta f_1 -\dfrac{\gamma}{m}(v-\omega)\dep{f_1}{v} \\
&+\dfrac{K}{2i m}\left( e^{i\alpha}e^{-i\theta}r_{1}[f_1\partial_v f^0]-e^{-i\alpha}e^{i\theta}r_{-1}[f_1\partial_v f^0] \right).
\end{split}
\end{equation}
$\L^\dagger$ is also diagonal when expressed in the Fourier basis with respect to $\theta$:
\[
\L^\dagger h = \sum_k (\L^\dagger_k h_k)e^{ik\theta}.
\] 
We thus concentrate on $\L^\dagger_1$, and look for $\tilde{\psi}$ such that
$\L^\dagger_1\tilde{\psi}=\lambda^\ast \tilde{\psi}$. Then
\begin{equation}
 (\lambda^\ast -iv)\tilde{\psi} +\frac{\gamma}{m}(v-\omega)\partial_v \tilde{\psi} = \frac{K}{2im}e^{-i\alpha} \int g(\omega). 
 \partial_v \tilde{\psi}(\omega,\omega)\,\d\omega
 \label{eq:tildevarpsi}
\end{equation}
It is not obvious how to compute $\tilde{\psi}$ from the above equation. Nevertheless, taking $v=\omega$, one easily extracts
$\tilde{\psi}(\omega,\omega)$:
\[
\tilde{\psi}(\omega,\omega) = \frac{K}{2im} e^{-i\alpha} \frac{C}{\lambda^\ast -i\omega}
\]
with $C$ a constant to be determined by normalization: we impose $\int \tilde{\psi}^\ast \psi \,\d v\,\d\omega =1$. 
Differentiating repeatedly \eqref{eq:tildevarpsi} with respect to $v$, and then taking $v=\omega$, one can compute
$\tilde{\psi}^{(n)}(\omega) = \partial_v^{n}[\tilde{\psi}](\omega,\omega)$ for any $n$. From \eqref{eq:varpsi}, we see that we need to compute up to $n=1$ 
in order to obtain $C$; the result is
\[
C = \frac{2im}{K}e^{i\alpha} \frac{1}{\Lambda'(\lambda)^\ast}.
\]
In the following, we will need
\begin{eqnarray}
\tilde{\psi}(\omega,\omega) &=& \frac{1}{\Lambda'(\lambda)^\ast}\frac{1}{\lambda^\ast -i\omega} \label{eq:psitilde0}\\
\tilde{\psi}^{(1)}(\omega) &=& \frac{i}{\Lambda'(\lambda)^\ast}\frac{1}{(\lambda^\ast -i\omega)(\lambda^\ast -i\omega+\gamma/m)} \label{eq:psitilde1}\\
\tilde{\psi}^{(2)}(\omega) &=& \frac{-2}{\Lambda'(\lambda)^\ast}\times \frac{1}{\Pi_{l=0}^2(\lambda^\ast -i\omega+l\gamma/m)}\label{eq:psitilde2}\\
\tilde{\psi}^{(3)}(\omega) &=& \frac{-6i}{\Lambda'(\lambda)^\ast}\times \frac{1}{\Pi_{l=0}^3(\lambda^\ast -i\omega+l\gamma/m)}\label{eq:psitilde3}
\end{eqnarray}
The eigenvectors of $\L^\dagger$ are then 
\[
\tilde{\Psi} = \frac{\tilde{\psi}(v,\omega)}{2\pi}e^{i\theta}~,~\tilde{\Psi}^\ast = \frac{\tilde{\psi}^\ast(v,\omega)}{2\pi}e^{-i\theta}~.
\]
The projection $\Pi$ onto the unstable eigenspace is defined by
\[
\Pi \cdot f = (\tilde{\Psi},f ) \Psi + (\tilde{\Psi}^\ast,f) \Psi^\ast
\]
and the orthogonal projection is $\Pi^\perp = 1-\Pi$.

\subsection{Unstable manifold}
We assume now that $\lambda_R=\Re(\lambda)>0$ and "small"; we will be in the end interested in the limit $\Re(\lambda)\to 0^+$.

The following computation essentially follows the steps of \cite{Crawford_Kura}. We want to describe the dynamics on the unstable manifold $\mathcal{M}$, 
which is two-dimensional and tangent to the unstable eigenspace $\mbox{Span}(\Psi,\Psi^\ast)$. 
We parameterize the unstable manifold by its projection onto the unstable eigenspace. Concretely, for $h\in \mathcal{M}$, we write
\[
h=f_0 + A \Psi +A^\ast \Psi^\ast +H[A,A^\ast](\theta,v,\omega)
\]
with $\Pi\cdot (h-f_0)= A\Psi+A\Psi^\ast$. Note that we assume that $H$ is a function of $A,A^\ast$; if
$\mathcal{M}$ "folds" on itself, $H$ may actually be multivalued. 

Writing as above $F=f_0+f$, with $f=A \Psi +A^\ast \Psi^\ast +S$, we have
\begin{eqnarray}
\dot{A}\Psi +\dot{A}^\ast \Psi^\ast + \partial_t S &=& \L\cdot (A \Psi +A^\ast \Psi^\ast +S) 
\\& &+ \mathcal{N}[A \Psi +A^\ast \Psi^\ast +S]. \nonumber
\end{eqnarray}
Applying the projections $\Pi$ and $1-\Pi$, we get
\begin{eqnarray}
\dot{A} &=& \lambda A + (\tilde{\Psi},\mathcal{N}[f])  \label{eq:A}\\
\partial_t S &=& \L S + \mathcal{N}[f] - \left[ (\tilde{\Psi},\mathcal{N}[f])\Psi +\mbox{c.c.}\right] \label{eq:S}
\end{eqnarray}
Using that $F\in \mathcal{M}$, we have $S=H(A,A^\ast)$, hence $\partial_t S=\dot{A}\partial_A H+\dot{A}^\ast \partial_{A^\ast} H$.
We introduce $(H_k)_{k\in \mathbb{Z}}$ the Fourier components of $H$:
\[
H = \sum_{k \in \mathbb{Z}} H_k(A,A^\ast,v,\omega) e^{ik\theta}.
\]
We write $\sigma=|A|^2$. To comply with the rotation symmetry $\mathcal{SO}(2)$, the Fourier components of $H$ have the following forms 
\begin{eqnarray}
H_0 &=& \sigma h_0(\sigma,v,\omega)   \label{eq:H0} \\ 
H_1 &=& A\sigma h_1(\sigma,v,\omega) \label{eq:H1}\\
H_k &=& A^k h_k(\sigma,v,\omega) \label{eq:Hk} 
\end{eqnarray}
with $H_{-k}=H_k^\ast$.
Note that the order parameter $r=|r_{1}|=|r_{-1}|$ is directly related to $A$, by $r_{-1}=2\pi A+O(A^3)$.

At the leading non linear order, only $h_0$ and $h_2$ contribute
to \eqref{eq:A}, which reads:
\begin{equation}
\dot{A} = \lambda A + A \sigma \dfrac{2\pi K}{2im} \left (e^{i\alpha} \langle \tilde{\psi}, \partial_v h_0\rangle - 
e^{-i\alpha}\langle \tilde{\psi}, \partial_v h_2\rangle\right ).
\label{eq:unstable_mani}
\end{equation}
We have used $\int \psi \,\d v\,\d\omega=\int \psi^\ast \,\d v\,\d\omega=1$ and the notation 
$\langle f_1,f_2\rangle = \iint f_1^\ast f_2 \,\d v\,\d\omega$ for the scalar product of two functions depending only on $(\omega,v)$.
We need now to compute $h_0$ and $h_2$ at leading order.
We introduce the expansions: $h_0(\sigma, v,\omega) = h_{0,0}(v,\omega) + O(\sigma)$ and $h_2(\sigma, v,\omega) = h_{2,0}(v,\omega) + O(\sigma)$.
From \eqref{eq:H0} and \eqref{eq:Hk}, we have
\begin{eqnarray} 
\frac{dH_0}{dt} &=& (A\dot{A}^\ast + A^\ast\dot{A})h_{0,0}+O(A^4)\nonumber\\
&=&2\lambda_r \sigma h_{0,0} +O(A^4) \nonumber \\
\frac{dH_2}{dt} &=& 2 \lambda A^2 h_{2,0}+O(A^4). \nonumber
\end{eqnarray}
From \eqref{eq:pde_nonlin} we get for the Fourier components of $\N[f]$:
\begin{eqnarray}
\left( \mathcal{N}[f] \right)_0 &=& i\dfrac{2\pi Ke^{-i\alpha}}{2m} \sigma\partial_v \psi  + \mbox{c.c.}+\mbox{higher~order} \nonumber \\
\left( \mathcal{N}[f] \right)_2 &=& -i\dfrac{2\pi Ke^{i\alpha}}{2m} A^2 \partial_v \psi  +\mbox{higher~order}. \nonumber
\end{eqnarray}
Thus, using \eqref{eq:S} we have to solve
\begin{eqnarray}
(2\lambda_R-\mathcal{L}_0)\cdot h_{0,0} &=& i\dfrac{2\pi Ke^{-i\alpha}}{2m} \partial_v \psi +\mbox{c.c.}  \label{eq:h00}\\
(2\lambda -\mathcal{L}_2)\cdot h_{2,0} &=& -i\dfrac{2\pi Ke^{i\alpha}}{2m}\partial_v \psi \label{eq:h20}.
\end{eqnarray}

\subsubsection{Computation of $h_{0,0}$}
We start from \eqref{eq:h00}. We have $h_{0,0}=h+\mbox{c.c.}$, where $h$ is the solution of
\begin{equation}
(2\lambda_R-\mathcal{L}_0)\cdot h= i\dfrac{2\pi Ke^{-i\alpha}}{2m}  \partial_v \psi. 
\label{eq:h00bis}
\end{equation}
\eqref{eq:h00bis} reads
\begin{equation}
\begin{split}
&2\lambda_R -\frac{\gamma}{m}\partial_v[(v-\omega) h_{0,0}] = \\
&\frac{2\pi K^2}{4im^2}\left(\frac{g\delta'(v-\omega)}{(\lambda^\ast-i\omega)(\lambda^\ast-i\omega+\gamma/m)} +i \frac{g\delta''(v-\omega)}{(\lambda^\ast-i\omega+\gamma/m)}\right)
\end{split}
\end{equation}
We introduce the ansatz:
\[
h = W_0(\omega)\delta(v-\omega) + W_1(\omega)\delta'(v-\omega)+ W_2(\omega)\delta''(v-\omega). 
\]
Using the identities
\begin{eqnarray*}
x\delta'(x) &=& -\delta(x) \\
x\delta''(x) &=& -2\delta'(x),
\end{eqnarray*}
we obtain
\begin{eqnarray}
W_0(\omega) &=& 0  \label{eq:W0}\\
W_1(\omega) &=& \frac{2 i\pi (K/2m)^2g(\omega)}{(2\lambda_R+\gamma /m)(\lambda+i\omega)(\lambda+\gamma /m+i\omega)} \label{eq:W1}\\
W_2(\omega) &=& \frac{2\pi (K/2m)^2g(\omega)}{2(\lambda_R+\gamma /m)(\lambda+\gamma /m+i\omega)}.\label{eq:W2}
\end{eqnarray}

\subsubsection{Computation of $h_{2,0}$}
A similar computation starting from \eqref{eq:h20} yields $h_{2,0}$.
We have to solve
\begin{equation}
(2\lambda -\mathcal{L}_2)\cdot h_{2,0} = -i\dfrac{2\pi Ke^{i\alpha}}{2m}\partial_v \psi.
\end{equation}
Using the ansatz
\[
h_{2,0} = X_0\delta(v-\omega) +X_1\delta'(v-\omega) +X_2\delta''(v-\omega), 
\]
we obtain
\begin{eqnarray}
X_0(\omega) &=& \frac{iX_1(\omega)}{(\lambda+i\omega)} \nonumber \\
X_1(\omega) &=& \frac{- i(2\pi Ke^{i\alpha}/2m)U_0(\omega)}{(2\lambda+2i\omega +\gamma /m)}+\frac{4iX_2(\omega)}{(2\lambda+2i\omega+\gamma /m)}\nonumber \\
X_2(\omega) &=& \frac{-i(2\pi Ke^{i\alpha}/2m) U_1(\omega)}{2(\lambda+i\omega+\gamma /m)}.\nonumber
\end{eqnarray}

\subsubsection{Putting everything together}

Inserting the expressions of $h_{0,0}$ and $h_{2,0}$ into \eqref{eq:unstable_mani}, we obtain the final reduced equation for $A$ we were looking for.
Let us start with the first contribution, which comes from $\langle \tilde{\psi}, \partial_v h_{0,0}\rangle$:
\begin{eqnarray}
\langle \tilde{\psi}, \partial_v h_{0,0} \rangle &=& 
\iint \tilde{\psi}^\ast (v,\omega) \left[ \left(W_1(\omega)+W_1^\ast(\omega)\right)\delta^{(2)}(v-\omega) \right. \nonumber\\
&& \left. +\left(W_2(\omega)+W_2^\ast(\omega)\right)\delta^{(3)}(v-\omega)\right] \,\d v\,\d\omega \nonumber \\
&=& \int \left[\tilde{\psi}^{(2)\ast}(\omega) \left(W_1(\omega)+W_1^\ast(\omega)\right)\right.\nonumber \\ 
&&\left.-\tilde{\psi}^{(3)\ast}(\omega) \left(W_2(\omega)+W_2^\ast(\omega)\right)\right]\,\d\omega.
\label{eq:psih00}
\end{eqnarray}
We have to compute the above integrals in the limit $\lambda_R \to 0^+$. A pole which moves to the real axis when $\lambda_R \to 0^+$ does not create any divergence by itself: although the integral is not well defined a priori, it can be analytically continued. However, divergences may appear through 
"pinching singularities", that is when two poles approach the real axis, each on one side. From \eqref{eq:psitilde1}, \eqref{eq:psitilde2}, \eqref{eq:psitilde3} and 
\eqref{eq:W1}, \eqref{eq:W2} one sees that a pinching singularity appears 
only in $\int \tilde{\psi}^{(2)\ast}W_1^\ast$; hence this provides the leading term:
\begin{equation}
\begin{split}
&\int \tilde{\psi}^{(2)\ast} W_1^\ast \,\d\omega  = i\pi \frac{K^2}{m^2}\frac{1}{(\gamma/m)}\frac{1}{\Lambda'(i\lambda_I)} \int \Bigg ( g(\omega) \times \nonumber \\
&\frac{1}{(\lambda_R^2 +(\omega+\lambda_I)^2)((\lambda_R+\gamma/m)^2+(\omega+\lambda_I)^2)}\times  \nonumber\\
&\frac{1}{(\lambda_R+2\gamma/m+i(\omega+\lambda_I))}\Bigg )\,\d\omega \nonumber\\
&\sim  i\pi \frac{K^2}{m^2}\frac{1}{(\gamma/m)^4}\frac{1}{\Lambda'(i\lambda_I)}\frac{\pi}{2}\frac{g(-\lambda_I)}{\lambda_R},
\end{split}
\end{equation}
where we have used
\begin{equation}
\int \frac{\varphi(x)}{x^2+\varepsilon^2} \underset{\varepsilon \to 0^+}{\sim} \pi \frac{\varphi(0)}{\varepsilon}.
\end{equation}

Let us turn to the second contribution, coming from $\langle \tilde{\psi}, \partial_v h_{2,0}\rangle$:
\begin{eqnarray}
\langle \tilde{\psi}, \partial_v h_{2,0} \rangle &=& \iint \tilde{\psi}^\ast (v,\omega) \Big[ X_0(\omega)\delta^{\prime}(v-\omega) \nonumber \\
&&\left.+X_1(\omega)\delta''(v-\omega)+X_2(\omega)\delta^{(3)}(v-\omega)\right]\d v\,\d\omega \nonumber\\
&=& \int \left[-\tilde{\psi}^{(1)\ast}(\omega) X_0(\omega)+\tilde{\psi}^{(2)\ast}(\omega) X_1(\omega) \right.\nonumber \\
&&\left.-\tilde{\psi}^{(3)\ast}(\omega) X_2(\omega)\right]\d\omega.  \label{eq:comph20}
\end{eqnarray}
It is not difficult to see that no pinching singularity appears, so that the above term has a finite limit when $\lambda_R\to 0$.

We conclude that the leading behavior of $c_3$ for $m>0$ is given by:
\begin{equation}
c_3 \sim \frac{\pi^3}{2}\frac{mK^3}{\gamma^4}\frac{e^{i\alpha}}{\Lambda'(i\lambda_I)}\frac{g(-\lambda_I)}{\lambda_R}. 
\label{eq:c3eq}
\end{equation}
In particular, the sign of $s(\alpha)={\rm Re}\left(\frac{e^{i\alpha}}{\Lambda'(i\lambda_I)}\right)$ determines the type (sub- or super-critical) of the bifurcation.

\subsection{Standard Kuramoto limit, $m\to 0$}
We have to take first the limit $m\to 0$, before $\lambda_R \to 0$. Counting the powers of $m$ in \eqref{eq:psih00} shows that
the whole contribution of $h_{0,0}$ vanishes in this limit, even taking into account the overall $1/m$ factor in front of the $O(A^3)$ term, see \eqref{eq:unstable_mani}.
Similarly, the $X_1$ and $X_2$ terms in \eqref{eq:comph20} give a vanishing contribution in the $m\to 0$ limit. Let us estimate the $X_0$ term:
\begin{equation}
 \int -\tilde{\psi}^{(1)\ast}(\omega) X_0(\omega)\,\d\omega \underset{m\to 0^+}{\sim} \frac{i m\, e^{2i\alpha}}{\Lambda'(\lambda)}\frac{\pi K^2}{2\gamma^3}\int \frac{g(\omega)}{(\lambda+i\omega)^3}\,\d\omega. \nonumber 
\end{equation}

One may then take the $\lambda_R\to 0^+$ limit, and this yields the following result
\begin{equation}
\lim _{\lambda_R \to 0^+}c_3 = \frac{\pi^2 K^2}{2\gamma^2} \frac{\Lambda''(i\lambda_I)}{\Lambda'(i\lambda_I)}, 
\end{equation}
where we have used the expression for $\Lambda$ in the limit $m\to 0$:
\[
\Lambda(\lambda) = 1-\frac{K}{2\gamma}e^{i\alpha}\int \frac{g(\omega)}{\lambda+i\omega}\,\d\omega.
\]
This recovers the expression for the standard Kuramoto model, see Eq.(139) in \cite{Crawford_Kura}.
In particular, the sign of $s^0(\alpha)={\rm Re}\left(\frac{\Lambda''(i\lambda_I)}{\Lambda'(i\lambda_I)}\right)$ determines the type of the bifurcation: 
$s^0>0$ (resp. $s^0<0$) corresponds to a subcritical (resp. supercritical) bifurcation.

\subsection{Hamiltonian (Vlasov) limit, $\alpha=0$, $\gamma \to 0$}
The Vlasov limit consists in taking $\gamma\to 0$ (cancelling the friction and the natural frequency driving), $\alpha=0$ (no shift between oscillator). As in the general case the $h_{2,0}$ term does not give any pinching singularity. Here as in \cite{Crawford_Vlasov} we use a fraction decomposition to get 
\begin{equation}
\begin{split}
&\int \left (\tilde{\psi}^{(2)\ast}(W_1^\ast+W_1)-\tilde{\psi}^{(3)\ast}(W_2^\ast+W_2)\right )\, \d \omega= 
\\
&\dfrac{-2i\pi}{\lambda_R\Lambda'(\lambda)}\left (\dfrac{K}{2m}\right )^2\int \dfrac{g(\omega)\left(3\lambda^\star-\lambda-4 i \omega\right )}{(\lambda+i \omega)^4(\lambda^\star-i \omega)^2} \,\d \omega+O\left (\lambda_R^{-1}\right )
\\
&=\dfrac{2i\pi}{\Lambda'(\lambda)}\left (\dfrac{K}{2m}\right )^2\int\Bigg (\frac{-1}{8 \lambda_R ^4 (\lambda+i\omega)^2}+\frac{1}{8 \lambda_R^4 (\lambda^\star-i\omega)^2}
\\&~~~~~~~~~~~~~~~~~~~~~~~~~~~~~~~~~-\frac{1}{2 \lambda_R ^3 (\lambda +i\omega  )^3}\Bigg )g(\omega)\d\omega+O(\lambda_R^{-2})
\\
&=\dfrac{i\pi}{\Lambda'(\lambda)}\dfrac{K}{2m}\left (\dfrac{\Lambda(\lambda)-1-\Lambda(\lambda^\ast)+1}{4\lambda_R^4}-\dfrac{\Lambda'(\lambda)}{2\lambda_R^3}\right )+O\left (\lambda_R^{-2}\right )\\
& = -\dfrac{i\pi}{4\lambda_R^3}\dfrac{K}{m}+O\left (\lambda_R^{-2}\right ),
\end{split}
\end{equation}
where we have used $\Lambda(\lambda)=\Lambda(\lambda^\ast)=0$.
Finally, in the limit $\lambda_R\to 0^+$
\begin{equation}
c_3 \sim -\frac{\pi^2 K^2}{4m^2} \frac{1}{\lambda_R^3}.
\end{equation}
As noted by Crawford this result does not depend on the initial velocity distribution. The $1/\lambda_R^3$ divergence yields the well know trapping scaling for the instability's saturation amplitude $|A|_{\infty}\propto \lambda_R^2$.

\section{The self-consistent mean-field method, and bistable behavior}

The self-consistent method (introduced in the original Kuramoto article \cite{Kuramoto75}, and later adapted to 
the case with inertia \cite{Tanaka97,Tanaka97b}) is a standard tool to understand qualitatively Kuramoto-like models.
We show here that:\\
i) The bistability of single oscillators pointed out in \cite{Tanaka97,Tanaka97b}  as the origin of the hysteretic behavior at large inertia cannot explain
the results at small mass presented in Fig.~1 of the main article.\\
ii) Nevertheless, the self-consistent method does predict a discontinuous transition for the parameters of Fig.~1, although it is difficult to make a general statement.

The basis of the method is to assume a constant value for the order parameter $r$. Then, considering the dynamics of the single oscillators
with this $r$, one may evaluate the contribution of each oscillator to the order parameter, and write a self consistent equation.

\subsection{Bistable behavior}
We assume that $r$ is fixed, and take its phase to be $0$ without loss of generality. The dynamics for a single oscillator with intrinsic frequency $\omega$ is
\begin{equation}
m\ddot{\theta} +\gamma \dot{\theta} = \gamma \omega -Kr\sin\theta.
\label{eq:base}
\end{equation}
Through the change of variable $t=Ts$ with $T=\gamma/Kr$, the dynamics reduces to (keeping the notation $\theta$)
\begin{equation}
\tilde{m}\frac{d^2\theta}{ds^2} +\frac{d\theta}{ds} = \tilde{\omega}-\sin \theta,
\label{eq:reduced_dyn}
\end{equation}
with $\tilde{m}=mKr/\gamma^2$ and $\tilde{\omega}=\gamma \omega/ Kr$. We have only two parameters.
When $\tilde{m}=0$, \eqref{eq:reduced_dyn} has a single attractive fixed point for small $\tilde{\omega}$ (corresponding to phase locked oscillators);
this fixed point collides with an unstable one for $\tilde{\omega}=1$, and the dynamics becomes periodic (drifting oscillators).
This behavior persists for small enough $\tilde{m}$. However, a qualitative change occurs for $\tilde{m}=\tilde{m}_{crit}\simeq 0.83$. Beyond this point, 
there is a range of values for $\tilde{\omega}$ where the stable fixed point coexists with an attractive periodic orbit: the dynamics \eqref{eq:reduced_dyn} is bistable.

Notice that the curves presented on Fig~1 of the main article for $m=0.25$ and $m=0.5$ feature in the transition region $K<3$ and $r<0.5$; thus the reduced mass 
$\tilde{m}<\tilde{m}_{crit}$, and bistability of the single oscillator dynamics cannot explain the discontinuous transition.

Nevertheless, it is possible that the self-consistent mean-field method predict a discontinuous transition, even without bistability of the single oscillator dynamics.

\subsection{Self-consistent equation}

We compute now the self-consistent equation "\`a la Kuramoto" for the parameters of Fig.~1, $m=0.5$; the starting point is \eqref{eq:base}. We have seen that there is no bistability for individual oscillators (at least in the transition region). Thus, the self-consistent equation simply reads as the sum of the contributions of locked and drifting oscillators:
\begin{equation}
r = r_{\rm locked} +r_{\rm drift}.
\label{eq:self}
\end{equation} 
The locked part is \cite{Tanaka97}
\begin{equation}
r_{\rm locked} =Kr \int_{-\pi/2}^{\pi/2} \cos^2\theta g(Kr\sin\theta)d\theta.
\label{eq:locked}
\end{equation}
The drifting part is more involved:
\begin{equation}
r_{\rm drift} = 2\int_{\omega>Kr} \frac{g(\omega)}{T(\omega)} \int_{-\pi}^{\pi} \frac{\cos \theta}{V_\omega(\theta)}d\theta,
\label{eq:drift}
\end{equation}
where $(\theta,V_\omega(\theta))$ is the attractive periodic orbit for an oscillator with intrinsic frequency $\omega$, and 
$T(\omega)$ is the period of this orbit. The factor $2$ in front comes from the $\omega \to -\omega$ symmetry.
\eqref{eq:drift} is usually computed in the large $m$ regime (or rather large $\tilde{m}$), which is of no interest to us. It would be possible to 
perform a small $\tilde{m}$ expansion. The results presented on Fig.~1 rely instead on a direct numerical estimation of \eqref{eq:locked} and 
\eqref{eq:drift}. For small $K$, \eqref{eq:self} has a single solution, $r=0$. Increasing $K$, two new solutions appear $r_<$ and $r_>$, at finite 
distance from $0$. On Fig.~1, we have plotted the $r_>$ solution as soon as it appears, although the $r=0$ solution may still be stable. 
We see that this self consistent method i) does predict a discontinuous transition for these parameters, and is in fair quantitative agreement 
with the numerical data ii) does not easily provide general statements about the transition, for different values of the parameters, and different 
frequency distributions.


\begin{thebibliography}{99}
\bibitem{Pikovsky} A. Pikovsky, M. Rosenblum and J. Kurths, "Synchronization: A universal concept in nonlinear sciences", Cambridge University Press, 2001.
\bibitem{Winfree} A. T. Winfree, "Biological rhythms and the behavior of populations of coupled oscillators", \emph{J. Theor. Biol.} {\bf 16}, 15 (1967).
\bibitem{Kuramoto75} Kuramoto, Yoshiki, H. Araki, ed. Lecture Notes in Physics, International Symposium on Mathematical Problems in Theoretical Physics 39. Springer-Verlag, New York. p. 420 (1975).
\bibitem{Daido92} H. Daido "Order Function and Macroscopic Mutual Entrainment in Uniformly Coupled Limit-Cycle Oscillators" \emph{Prog. Theor. Phys.} {\bf 88}, 1213-1218 (1992).
\bibitem{Sakaguchi88} H. Sakaguchi "Cooperative Phenomena in Coupled Oscillator Systems under External Fields" \emph{Prog. Theor. Phys.} {\bf 79}, 39-46 (1988). 
\bibitem{Sakaguchi88b} H. Sakaguchi, S. Shinomoto and Y. Kuramoto "Mutual Entrainment in Oscillator Lattices with Nonvariational Type Interaction" \emph{Prog. Theor. Phys.}{\bf 79}, 1069-1079 (1988).
\bibitem{Izhikevich98} E.M. Izhikevich "Phase models with explicit time delays" \emph{Phys. Rev. E} {\bf 58}, 905 (1998).
\bibitem{Yeung99} M.K.S. Yeung and S.H. Strogatz "Time Delay in the Kuramoto Model of Coupled Oscillators"
\emph{Phys. Rev. Lett.}{\bf 82}, 648 (1999).
\bibitem{Sakaguchi87} H. Sakaguchi, S. Shinomoto, and Y. Kuramoto, "Local and grobal self-entrainments in oscillator lattices", \emph{Prog. Theor. Phys.} {\bf 77}, 1005 (1987).
\bibitem{Hong02} H. Hong, M. Y. Choi, and Beom Jun Kim "Synchronization on small-world networks"
\emph{Phys. Rev. E} {\bf 65}, 026139 (2002).
\bibitem{Ermentrout91} B. Ermentrout "An adaptive model for synchrony in the firefly Pteroptyx malaccae" \emph{Journal of Mathematical Biology} {\bf 29}, 571-585 (1991).
\bibitem{Wiesenfeld96} K. Wiesenfeld, P. Colet, S.H. Strogatz, "Synchronization transitions in a disordered Josephson series array", \emph{Phys. Rev. Lett.} {\bf 76}, 404 (1996).
\bibitem{Trees05} B. Trees, V. Saranathan and D. Stroud "Synchronization in disordered Josephson junction arrays:
Small-world connections and the Kuramoto model" \emph{Phys. Rev. E} {\bf 71}, 016215 (2005).
\bibitem{Filatrella08} G. Filatrella, A.H. Nielsen and N.F. Pedersen "Analysis of a power grid using a Kuramoto-like model" \emph{Euro. Phys. Jour. B} {\bf 61}, 485-491 (2008).
\bibitem{Olmi14} S. Olmi, A. Navas, S. Boccaletti, and A. Torcini "Hysteretic transitions in the Kuramoto model with inertia" \emph{Phys. Rev. E} {\bf 90}, 042905 (2014).
\bibitem{Ji13} P. Ji, T.K.D. Peron, P.J. Menck, F.A. Rodrigues and J. Kurths "Cluster explosive synchronization in complex networks" \emph{Phys. Rev. Lett.} {\bf 110}, 218701 (2013).
\bibitem{Tanaka97} H.-A. Tanaka, A. J. Lichtenberg, and S. Oishi, "First order phase transition resulting from finite inertia in coupled oscillator systems", \emph{Phys. Rev. Lett.} {\bf 78}, 2104 (1997).
\bibitem{Acebron98} J. A. Acebr\'on and R. Spigler "Adaptive Frequency Model for Phase-Frequency Synchronization in Large Populations of Globally Coupled Nonlinear Oscillators" \emph{Phys. Rev. Lett.}{\bf 81}, 2229 (1998).
\bibitem{Tanaka97b} H.-A. Tanaka, A. J. Lichtenberg, and S. Oishi, "Self-synchronization of coupled oscillators with hysteretic responses", \emph{Physica D} {\bf 100}, 279
(1997).
\bibitem{Crawford_Vlasov} J.D. Crawford, "Amplitude equations for electrostatic waves: universal singular behavior in the limit of weak instability" \emph{Physics of Plasmas} 
{\bf 2}, 97-128 (1995).
\bibitem{Crawford_Kura} J.D. Crawford, "Amplitude expansions for instabilities in populations of globally-coupled oscillators", \emph{J. Stat. Phys.}
March 1994, Volume 74, Issue 5, pp 1047-1084
\bibitem{Crawford_Kura_prl} J.D. Crawford, "Scaling and Singularities in the Entrainment of Globally Coupled Oscillators", \emph{Phys. Rev. Lett.} {\bf 74}, 4341 (1995).
\bibitem{Strogatz_review} S.H. Strogatz, "From Kuramoto to Crawford: Exploring the onset of synchronization in populations of coupled oscillators" \emph{Physica D} {\bf 143}, 1-20 (2000).
\bibitem{Ott08} E. Ott and T.M. Antonsen, "Low dimensional behavior of large systems of globally coupled oscillators"
\emph{Chaos} {\bf 18}, 037113 (2008).
\bibitem{Martens09} E. A. Martens, E. Barreto, S. H. Strogatz, E. Ott, P. So, and T. M. Antonsen, "Exact results for the Kuramoto model with a bimodal frequency distribution" \emph{Phys. Rev. E} {\bf 79}, 026204 (2009).
\bibitem{Acebron00} J. A. Acebr\'on, L. L. Bonilla, and R. Spigler, "Synchronization in populations of globally coupled oscillators with inertial effects"
\emph{Phys. Rev. E} {\bf 62}, 3437 (2000).
\bibitem{Acebron05} J.A. Acebr\'on, L.L. Bonilla, C.J. Vicente-P\'erez, F. Ritort, R. Spigler, "The Kuramoto model: A simple paradigm for synchronization phenomena", \emph{Reviews of Modern Physics} {\bf 77},137-185 (2005).
\bibitem{Komarov14} M. Komarov, S. Gupta and A. Pikovsky "Synchronization transitions in globally coupled rotors in the presence of noise and inertia: Exact results", \emph{Europhysics Letters}{\bf 106}, 40003 (2014). 
\bibitem{Campa09} A. Campa, T. Dauxois and S. Ruffo, 
"Statistical mechanics and dynamics of solvable models with long-range interactions",
\emph{Phys. Rep.} {\bf 480}, 57 (2009).
\bibitem{suppl} See supplemental material for the details.
\bibitem{Omel12} O. E. Omel'chenko and M. Wolfrum, "Nonuniversal Transitions to Synchrony in the Sakaguchi-Kuramoto Model" \emph{Phys. Rev. Lett.} {\bf 109}, 164101 (2012).
\bibitem{Gupta14} S. Gupta, A. Campa, and S. Ruffo ,"Nonequilibrium first-order phase transition in coupled oscillator systems with inertia and noise",
\emph{Phys. Rev. E} {\bf 89}, 022123 (2014).
\bibitem{Hong07} H. Hong, H. Chat\'e, H. Park, and L.-H. Tang "Entrainment Transition in Populations of Random Frequency Oscillators"
\emph{Phys. Rev. Lett.} {\bf 99}, 184101 (2007).
\bibitem{Chiba} H. Chiba, "A proof of the Kuramoto conjecture for a bifurcation structure of the infinite-dimensional Kuramoto model", \emph{Ergodic Theory and Dynamical Systems}, {\bf 35}, 762-834 (2015).
\bibitem{Dietert} H. Dietert, "Stability and bifurcation for the Kuramoto model", \emph{Journal de Math\'ematiques Pures et Appliqu\'ees} {\bf 105}, 451-489 (2015).


\end{thebibliography}
\end{document}